# An Efficient Cluster-based Routing Protocol in Cognitive Radio Network

Fatima Zohra Benidris[1], Badr Benmammar[1], Leila Merghem-Boulahia[2] and Moez Esseghir[2]

[1] LTT Laboratory, University of Tlemcen, Algeria
[2] ICD/ERA, Université de Technologie de Troyes, France
{fatima.benidriss, badr.benmammar}@gmail.com, {leila.merghem_boulahia, moez.esseghir}@utt.fr

***Abstract***: Cognitive Radio Networks (CRNs) are being studied intensively and gaining importance as spectrum is the heavily underutilized. CRN has the capability to exploit smartly the unutilized frequency spectrum. Recently, the research community started to work in the area of cognitive radio routing. In a flat topology, all nodes are of the same level and functionality, thus making it simple and efficient for smaller networks. However, when the network is large with sparse nodes, the routing information becomes more complex making cluster-based techniques really relevant to tackle such situations. In a cluster-based routing, all nodes in the network are dynamically organized into partitions called groups or clusters. In each cluster, a cluster head is chosen to help in the data transmission management and to maintain cluster membership information. This paper proposes a novel routing protocol for cognitive radio ad hoc networks (CRAHNs) based on clustering model which amends swiftly to the topological changes and establishes the routing efficiently. Our proposed approach is thoroughly evaluated through simulation study. The results state the suitability of the proposed protocol for cognitive radio ad hoc networks and demonstrate that it has better performance in terms of finding the source-destination route, reducing the amount of messages that are transmitted all over the network and minimizing the routing delay.

## 1. INTRODUCTION

Cognitive Radio is an emerging and promising technology that aims to increase the overall utilization of radio resources by enabling the dynamic allocation of some portions of the wireless spectrum. Unlicensed users, through cognitive radio devices, can opportunistically operate over the current unused parts of licensed bands called white spaces, spectrum holes, or spectrum opportunity [1].

The network mainly consists of two types of users: licensed and unlicensed users.

Licensed users are also known as primary users (PU) and unlicensed users are known as secondary user (SU). SU's access spectrum conditionally which means when primary users are inactive [2].

Although more than ten years have passed, the research on cognitive radio networks has mainly focused on physical and medium access issues [3, 4], including the definition of effective spectrum sensing, decision and sharing mechanisms. However, recently the research community has started working on cognitive radio routing.

The routing protocols determine how messages can be forwarded, from a source node to a destination node which is out of the range of the former, using other mobile nodes of the network. Routing, which includes for example maintenance and discovery of routes, is one of the very challenging areas in communication [5].

In this paper, we are interested in the routing problem in a cognitive radio ad-hoc network by proposing a novel routing protocol based on clustering mechanism which can take advantages of both reactive and proactive protocols. In a cluster-based routing, all nodes in the network are dynamically organized into partitions called groups or clusters. These clusters are then combined into larger partitions to help maintain a relatively stable network topology. The nodes that have been elected as cluster heads have the necessary intelligence to organize route forwarding and route maintenance procedure.

The rest of the paper is organized as follows. Section 2 describes the related works in this area. Section 3 presents the proposed approach and the results of an extensive performance evaluation are provided in Section 4. Finally, in Section 5 some conclusions and future works are drawn.

## 2. RELATED WORKS

A number of routing protocols have been proposed and implemented for CRAHNs in order to enhance the bandwidth utilization and provide higher throughputs, less overheads per packet and minimum consumption of energy. All these protocols have their own advantages and disadvantages under certain circumstances.

The papers [6, 7] propose to enhance the AODV protocol for cognitive radio scenarios. In [6], the authors propose to use a dedicated common control channel for route formation. Therefore, the data packets are routed along channels whose quality has not been assessed. In [7], the authors do not resort to a dedicated channel but, since they assume that nodes are equipped with a single transceiver, a similar issue due to the use of un-assessed channels arises.

In [8], a protocol for mobile CRAHNs based on the geographic forwarding paradigm has been proposed. The main idea of the protocol is to discover several paths, which are combined at the destination to form the path with the minimum hop count. This protocol it is able to deal with reasonable levels of PU activity changing rate. However, it assumes that most of the nodes are GPS equipped and, most importantly, a mechanism for disseminating the destination location both at the source and at each intermediate node is required.

Furthermore, in [9], the authors make use of communication locality and propose a new routing scheme, called proactive route maintenance. Routing information is disseminated along active routes and advertised by active nodes on the routes. Alternative paths are dynamically discovered and maintained by active nodes and their one-hop neighbors. In this way, they achieve the high delivery ratio, low latency and fair load distribution.

SAMER is a routing scheme proposed in [10] to provide a tradeoff between long-term parameters such as hop count and short-term parameters such as spectrum availability. Besides, SAMER provides a compromise between the local spectrum conditions at the forwarding nodes and the global spectrum view of the entire routing path. However, the spectrum availability does not appear well in the calculation of routing metrics except by the percentage of time during which channels are available for CR nodes. According to [11], the next forwarding node is the node that requires the minimal consumption power. Hence, the next forwarding node is usually the nearest neighboring node; and then the optimal route is much longer than the route of minimum hop count. Moreover, the percentage of channels availability is not taken into consideration during the channel selection.

In [12], a new routing metric was proposed based on a probabilistic definition of the available capacity over channels. This definition aims at finding the most probable route or the route with the higher probability of availability. After the route assignment, the source examines if its throughput capacity satisfies the throughput demand and it adds, when needed, other channels for transmission until the throughput demand is satisfied. Probabilistic throughput computation is adequate to increase the long term availability and the overall statistical utilization of the network. However, this approach may not be adapted for short connections.

## 3. THE PROPOSED APPROACH

In this section, we consider the architecture where the SUs and PUs are organize into clusters as shown in figure 1and these clusters cooperate with each other to make spectrum sharing which alleviates the spectrum scarcity problem.

In the cognitive radio network context, grouping nodes that share locally available spectrum holes enables nodes to efficiently coordinate their local interactions and improves the efficiency of network functions such as spectrum management, data transmission and energy consumption balance between CR users. One CR node is elected as a cluster-head in each cluster (based on one or more criteria such the powerful node in term of energy and has a long lifetime). The cluster-head helps in the data transmission management and maintains cluster membership information.

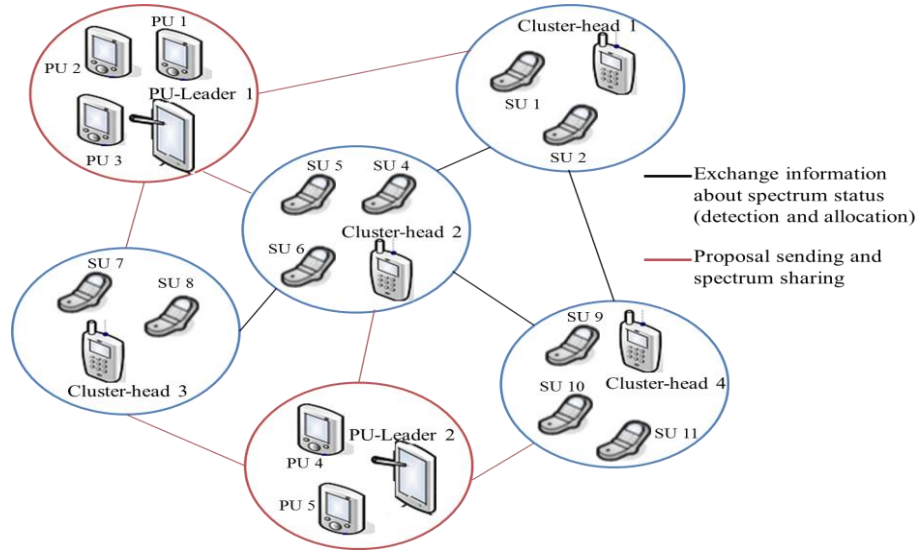

**Figure 1.** Cluster-based Cognitive radio network [15]

### 3.1 Cluster formation

Clustering is the process of partitioning or grouping a given set of patterns into disjoint clusters. K-means [13] is one of the simplest unsupervised learning algorithms that solve the well-known clustering problem. The CR node moves randomly in the network, so we adopt the K-Nearest neighbors [14] to determine to which cluster it belongs. When the CR node enters into the network it is in undecided state, it broadcasts hello messages to its neighbor nodes. When the hello messages are received by the neighbor node, it sends a triggered hello message to undecided node. The CR node use the delay of triggered hello message to calculates the indirect distance with this node. The undecided CR node selects K nodes with minimum delay, then it can decide at which cluster it belongs.

### 3.2 Cluster Head Selection

It is important for each cluster to have a cluster head to maintain and control the nodes' mobility and the routing decisions. Cluster head is also responsible for guaranteeing that none of its nodes in the cluster is overly burdened, as it might result in high cost and routing overhead (as a result of node movement, node failure or packet loss).

The cluster-head is chosen according to certain criteria, for example it is a static node or has energy resources superior to other nodes in the network, etc. Once a node is elected as a cluster head, it is desirable that the node remains the head up to a specified maximum amount of time to take into account wireless networks characteristics such as battery life of individual nodes, and location of the cluster head node within the cluster. The cluster head would also have the intelligence to organize route forwarding and route maintenance procedures. When a cluster head is about to 'resign or retire', it will send out a 'check_for_new_cluster_head' message to its neighbors in order to find a new head.

### 3.3 Cluster-based routing protocol

Having studied the related work done on routing protocol in cognitive radio network and taken into account the drawbacks and deficiencies of other approaches, the proposed cluster based routing protocol tackles these issues in terms of neighborhood discovery and route reconstruction/ maintenance phase. Our protocol can take advantages of both reactive and proactive protocols. When a source node S wants to transmit a message to destination and the route is unknown, S initiates a connection to D by performing Route Discovery. Then, the destination D performs Route Selection by choosing the optimal route based on potential routing metrics.

In our design, we assume that each cluster head maintains a neighbor table which contains the details of other cluster heads within its neighborhood. Also contains the details of all its cluster members in its routing table. The information of cluster members and neighbor cluster heads are obtained by exchanging the HELLO message.

In the following, we describe the components of our protocol in further detail.

### 3.3.1 Route discovery

When a route is needed and it is not already known by a node it sends a Route Request (RREQ) message to its cluster-head (Cs). Each RREQ message is identified by a request ID and the source and destination IP addresses. This is done in order to prevent that an intermediate node which receives twice the same RREQ message forwards it twice. The Cs broadcasts a discovery packet RREQ to its neighbors. Those forward the message until it reaches the destination node. If an additional request for the same pair source-destination is received by an intermediate cluster-head (Ci) it is simply discarded. As Ci receives a RREQ message, it first updates the RREQ with its address and examines the current partial path for the address of its members. If found, Ci sends Route Reply (RREP), and now drops RREQ message in order to break the routing loop. Otherwise, it looks in the routing table if it already knows the route to the destination, then the RREQ will not broadcast to neighboring cluster-heads but sent only to the cluster-head belonging to the path of the destination and if it does not know the path to the destination, it continues to broadcast the RREQ to neighboring cluster-heads until you find the destination. However, the routing is done in reactive manner if the path to the destination is not known and in a proactive manner if a partial path of the route is already known by a certain CHI (the following the path to the destination is already registered in its routing table). To avoid broadcast congestion in the CR network, the hops-count of routing discovery is limited by Hmax. So if the number of hope is greater that Hmax, the RREQ message is dropped to break the routing loop. Our protocol may discover multiple diverse paths to the destination. All possible paths are sent to the destination for route selection. When the cluster-head of the destination node (Cd) receives the first RREQ, it starts a timer TR and collects all RREQ messages until TR expires. When TR expires, Cd selects the optimal route and sends a RouteReply (RREP) back to the source. We illustrate the steps in RREQ processing in figure 2.

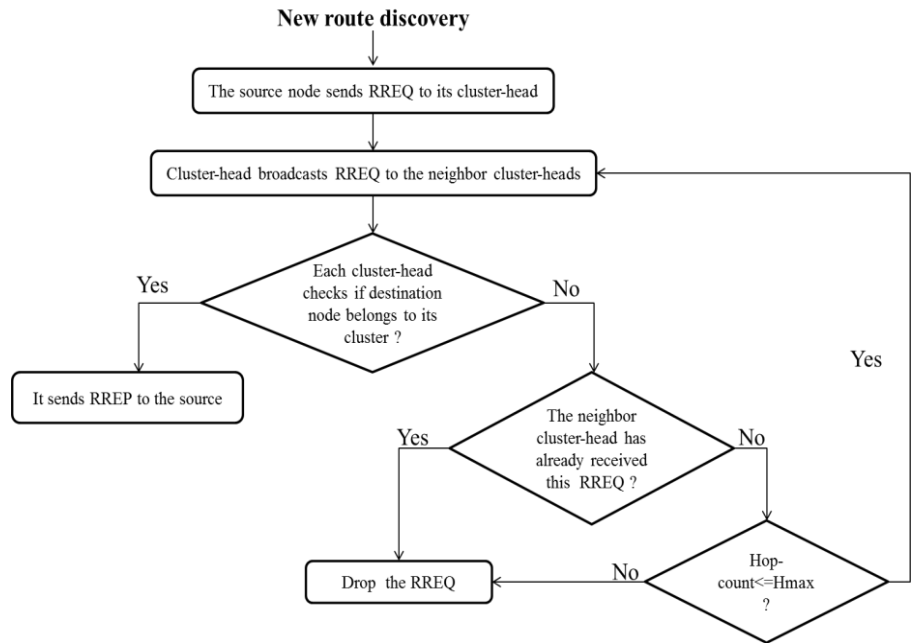

**Figure 2.** Route request flow chart

### 3.3.2 Route selection (Route formation)

In this section, we describe in detail how destination D analyzes its set of received paths to select the best route. The cluster-head of destination D can use a variety of policies to determine the most desirable route. Cd can estimate the route's maximum throughput. Cd can also seek to minimize end-to-end latency by choosing routes with lower hop-count. Finally, we can also consider quality-based metrics such as ETX (Expected Transmission Count) [15], ETT (Expected Transmission Time), WCETT (Weighted Cumulative ETT) [16]. In general, the Cd computes a policy-driven utility value for each candidate route based on a combination of the above factors, and selects the route with the highest utility. In our implementation, we select routes by using minimal hop-count (minimum number of intermediate cluster-head) and each intermediate cluster-head selects a CR node from its cluster that shares a common channel with the previous hop node and which has a maximum throughput. When the destination receives the RREQ it sends a Route Reply (RREP) message back to the source node along the optimal route by reversing the hop sequence recorded in the RREQ (the IDs of the cluster-head nodes). Each intermediate cluster-head updates the RREP with address of a selected CR node which has a maximum throughput. When the source receives the RREP it can start communicating with the destination.

### 3.3.3 Route maintenance

Topology changes due to node mobility or wireless propagation instability are handled with traditional route error (RERR) packets. When a route has been established, it is being maintained by the source node as long as the route is needed. Movements of nodes affect only the routes passing through this specific node and thus do not have global effects. When the source node moves while having an active session and loses connectivity with the next hop of the route; this source node sends RERR to its new cluster-head to establish a new route, if the new cluster-head belongs to the old way is not necessary to make a new route discovery, simply update the old route by removing the part that links the ancient cluster-head with the new cluster-head of the source. Otherwise, a new route discovery session is started by using a traditional RERR packet. If though an intermediate station loses connectivity with its next hop it initiates a Route Error (RERR) message and broadcasts it to its cluster-head, this later rebroadcasts the RERR to the cluster-head of failed node to select another intermediate node. Therefore, the message can be transmitted towards destination.

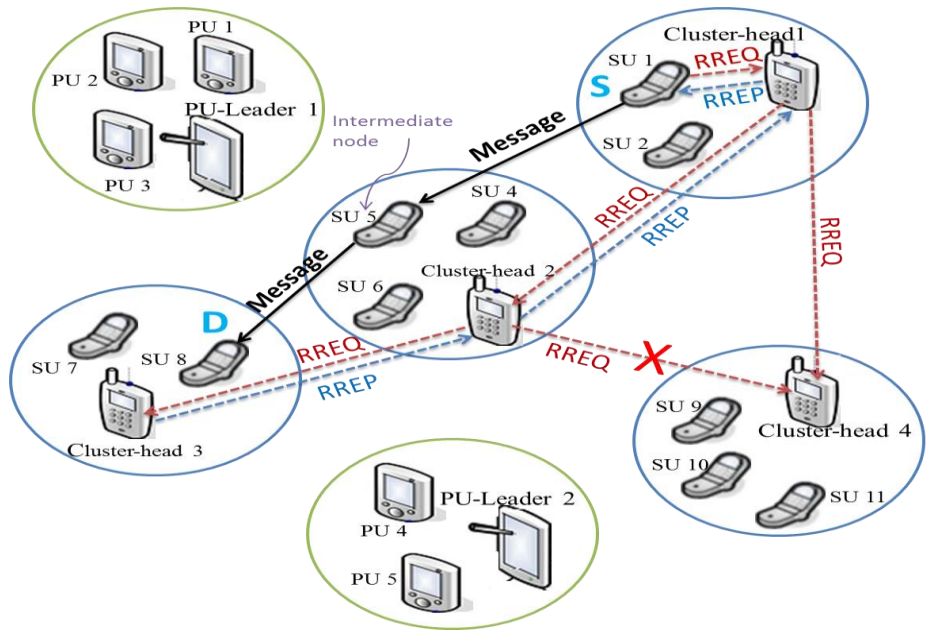

**Figure 3.** Establishing a multi-hop virtual circuit between source S and destination D using the proposed protocol.

### 3.3.4 Illustrative example

We illustrate the high level operation of the proposed protocol using an example. As shown in figure 3, node SU1 wants to send a message to destination SU8.Therefore, SU1 sends RREQ to cluster-head 1, this later broadcast the RREQ to the neighboring cluster-heads: cluster-head 2 and cluster-head 4. Cluster-head 2 broadcasts the RREQ to its neighbor cluster-heads: cluster-head 3 and cluster-head 4, the RREQ is dropped by cluster-head 4 because it's already received. As node cluster-head 3 receives a RREQ message, the destination SU8 is found in its cluster then the RREP is sent back to the source and each cluster-head appends the identifier of the most appropriate node to be intermediate node, in our example there is one intermediate node SU5 selected by cluster-head 2. Finally the source can forward the message to the destination along the following path {SU1, SU5, SU8}.

## 4. SIMULATION

If In this section, we present various numerical results to evaluate the working of the proposed protocol, based on the following simulation setup. We perform our simulations under the assumption of cognitive radio network where the secondary users are grouped into clusters and each cluster contains a cluster-head. We randomly distribute nodes in a $1000 \times 1000$ unit square. Some of the nodes are busy at a certain channel and some of the nodes are idle. We randomly choose a source and a destination. In our simulation we consider that the neighboring cluster-heads have a common channel to communicate between them.

Considering the capacity of a single machine, the maximum number of primary and secondary users is 100 in total. All the simulations are conducted in Java, overall PC with 2.4GHZ dual processor and 3GB memory.

We demonstrate the performance improvement attained by our protocol by comparative study of AODV (reactive routing protocol). Firstly, we are interested by comparing the number of message of both RREQ and RREP, then the routing delay and finally the successful rates of finding the source-destination route. For each scenario, ten runs with different random seeds were conducted and the results were averaged.

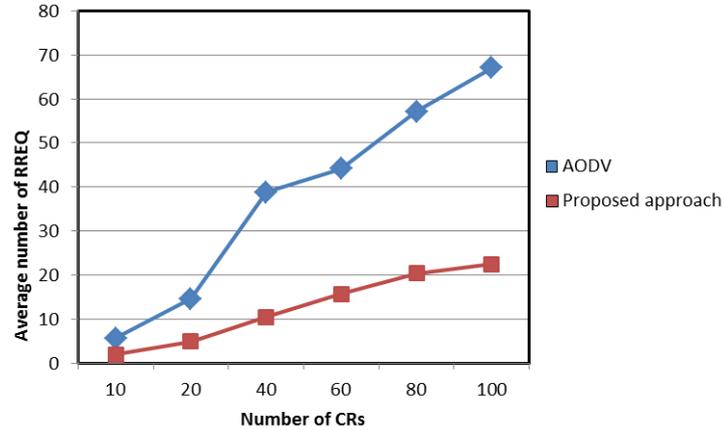

**Figure 4.** Average Number of RREQ with Number of CRs

Figure 4 depicts the average number of RREQ with different number of CRs. We observe that the number of RREQ is very high for AODV protocol and it is much larger than our approach when the number of CRs is between 40 and 100, because when a route is needed and it is not already known by a node it sends a RREQ message to its neighbors. Those forward the message until it reaches the destination node. So the number of RREQ increase quickly with increasing the number of CRs in the network unlike our approach where the number of RREQ increase slowly as showing in the figure because the users are grouped into clusters, which is known to be effective in reducing communication overhead in variable network environment showing the communication efficiency.

Figure 5 depicts the average number of RREP with different number of CRs. We notice that the proposed approach has a reduced number of requests comparing with AODV protocol which means that the number of hops is small in our approach because the RREP message back to the source through the cluster-heads of intermediate clusters which are recorded in the RREQ.

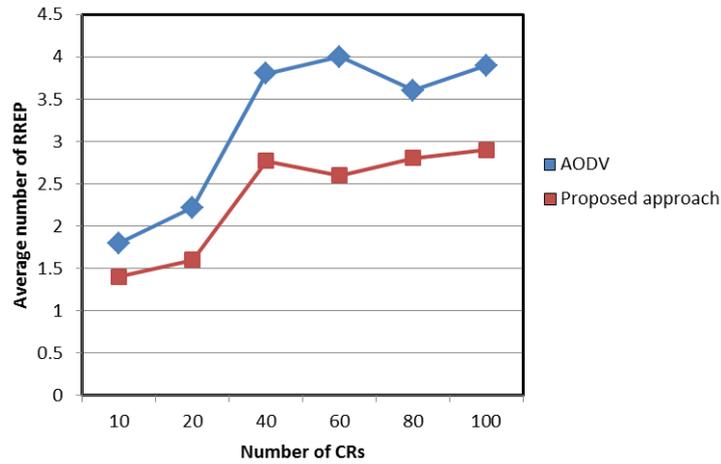

**Figure 5.** Average Number of RREP with Number of CRs

Figure 6 illustrates the average routing delay with several sets of CRs. The routing delay is time needed to find the destination. We note that the increasing pattern in routing delay is directly relational to the number of CRs and comparing our approach with AODV protocol, we have a better results in term of minimizing the routing delay because in AODV, the RREQ is broadcasted to all neighboring CR nodes until finding the destination which increase the routing delay but in our approach the clustering facilities the route discovery where the cluster-head has the information of its CRs members then it can know if the destination is in its cluster else it broadcast the request to the neighboring cluster head which minimize the routing delay.

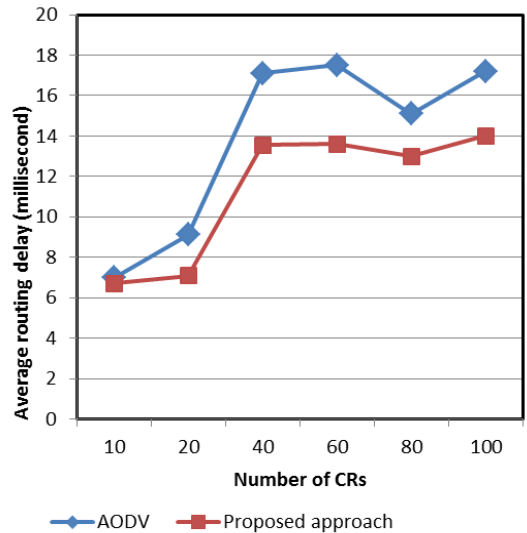

**Figure 6.** Average routing delay with Number of CRs

Figure 7 depicts the success rate histogram of finding the source-destination route. We note that the success rate is influenced by the number of SU in the network and it slowly decreases with the increase in the number of SU. We observe that the AODV approach has fewer successful rates than our approach where almost CRs are fully satisfied by finding the route to the destination, and the rate difference between the two protocols increases with increasing the number of SU. However, our protocol ensures the proper functioning of routing through clustering that facilitates the establishment of a route between the source and the destination.

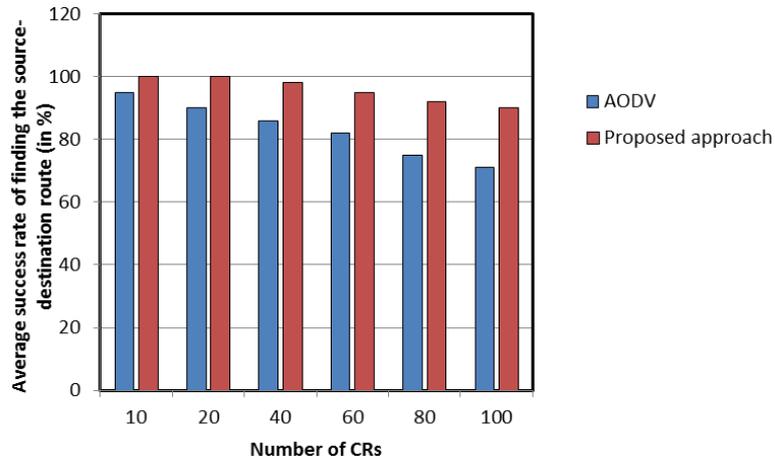

**Figure 7.** Success rate of finding the source-destination route.

## 5. CONCLUSIONS

This paper proposes a new routing approach in cognitive radio ad-hoc network based on clustering mechanism where the cluster-head is responsible of discovery and selection route. Our proposed protocol computes the route to the destination and then checks if this route is able to satisfy the required type of connection at the source.

Experimental results show that compared to AODV protocol, our protocol works very effectively by introducing cluster-head in each cluster in the network where the successful finding source-destination route is higher even with large number of CR nodes. Furthermore, our solution works very effectively without having higher routing delay and communication cost (RREQ and RREP messages).

In the future works, we will interest by studying the energy conservation in mobile cognitive radio network using our cluster based routing protocol mechanism to reduce loading of networks, energy conservation and increase lifetime of nodes and networks.